\documentclass[%
 aps,
 prl,%
 amsmath,amssymb,
 reprint,%
runinaddress,
showpacs,
superscriptaddress,
floatfix
]{revtex4-1}

\usepackage{graphicx}
\usepackage{epsfig}%
\usepackage{dcolumn}
\usepackage{bm}
\usepackage{multirow}
\usepackage{notoccite}

\begin{document}

\title{Nonlinear stabilization of tokamak microturbulence by fast ions}

\author{J.~Citrin} 
\affiliation{\mbox{Dutch Institute for Fundamental Energy Research DIFFER, Association EURATOM-FOM, Nieuwegein, The Netherlands}}
\author{C.~Bourdelle} 
\affiliation{CEA, IRFM, F-13108 Saint Paul Lez Durance, France}
\author{J.~Garcia} 
\affiliation{CEA, IRFM, F-13108 Saint Paul Lez Durance, France}
\author{J.W.~Haverkort} 
\affiliation{Centrum Wiskunde \& Informatica (CWI), PO Box 94079, Amsterdam, The Netherlands}
\affiliation{\mbox{Dutch Institute for Fundamental Energy Research DIFFER, Association EURATOM-FOM, Nieuwegein, The Netherlands}}
\author{G.M.D.~Hogeweij} 
\affiliation{\mbox{Dutch Institute for Fundamental Energy Research DIFFER, Association EURATOM-FOM, Nieuwegein, The Netherlands}}
\author{F.~Jenko} 
\affiliation{\mbox{Max Planck Institute for Plasma Physics, EURATOM Association, 85748 Garching, Germany}}
\author{T. Johnson} 
\affiliation{Euratom-VR Association, EES, KTH, Stockholm, Sweden}
\author{P.~Mantica} 
\affiliation{\mbox{Istituto di Fisica del Plasma ''P. Caldirola,'' Associazione Euratom-ENEA-CNR, Milano, Italy}}
\author{M.J.~Pueschel} 
\affiliation{University of Wisconsin-Madison, Madison, Wisconsin 53706, USA}
\author{D.~Told} 
\affiliation{\mbox{Max Planck Institute for Plasma Physics, EURATOM Association, 85748 Garching, Germany}}
\author{JET-EFDA contributors}
\affiliation{\mbox{See the Appendix of F. Romanelli et al., Proc.~of the 24th IAEA Fusion Energy Conference 2012, San Diego, USA}}
\collaboration{JET-EFDA, Culham Science Centre, Abingdon, OX14 3DB, UK}
\noaffiliation

\begin{abstract}
Nonlinear electromagnetic stabilization by suprathermal pressure gradients found in specific regimes is shown to be a key factor in reducing tokamak microturbulence, augmenting significantly the thermal pressure electromagnetic stabilization. Based on nonlinear gyrokinetic simulations investigating a set of ion heat transport experiments on the JET tokamak, described by Mantica \textit{et al.}~[Phys. Rev. Lett. \textbf{107} 135004 (2011)], this result explains the experimentally observed ion heat flux and stiffness reduction. These findings are expected to improve the extrapolation of advanced tokamak scenarios to reactor relevant regimes.
\end{abstract}


\maketitle

\textit{Introduction}.--It is well established that the primary limitation of energy confinement in tokamaks is turbulent transport driven by microinstabilities~\cite{ITER2}. The ion-temperature-gradient (ITG) instability~\cite{roma89}, in particular, has been long identified as an important driver of microturbulence, and is primarily responsible for ion heat losses. ITG modes are driven linearly unstable by logarithmic ion temperature gradients above a critical threshold, i.e., by $R/L_{Ti}>R/L_{Ti,\mathrm{crit}}$, where the tokamak major radius $R$ is a normalizing factor. The modes saturate in conjunction with nonlinearly excited zonal flows, forming a self-organized turbulent system which sets the transport fluxes~\cite{diam05}. In the following, we term `stiffness' the degree of sensitivity of the ion heat flux to the driving $R/L_{Ti}$. At lower stiffness, higher $R/L_{Ti}$ is attained for the same input heat flux and critical threshold.

Recent experiments have challenged our present theoretical understanding of ITG turbulence~\cite{mant09,mant11}. A significant reduction of ion stiffness was reported in conditions of concomitant low magnetic shear $\hat{s}$ and high rotational flow shear. However, until now, nonlinear gyrokinetic simulations have not reproduced the stiffness reduction, yielding absolute levels of ion heat flux well above the experimental measurements~\cite{mant11}. Understanding this phenomenon, and bridging the gap between the experimental observation and theoretical prediction, is critical for increased trust in extrapolations of turbulent transport to future devices such as ITER. 

In this Letter, we report on gyrokinetic simulations of a number of discharges from the data-set reported in Ref.~\cite{mant11}, using the \textsc{Gene} code~\cite{jenk00}. These simulations include for the first time modeled fast ion species arising from both neutral beam injection (NBI) and ion cyclotron resonance heating (ICRH), which provide suprathermal pressure. Nonlinear stabilization of ITG turbulence by both thermal and suprathermal pressure gradients is predicted to significantly reduce the simulated ion heat flux to levels in line with the measured values, explaining the observed stiffness reduction. This mechanism is shown to be more effective at low $\hat{s}$, in agreement with observations. 

Previously considered \textit{linear} mechanisms of fast ion stabilization of ITG modes are insufficient to explain the observed degree of flux reduction for these discharges. These include linear electromagnetic (i.e., including both electric \textit{and} magnetic field fluctuations in the model) stabilization by suprathermal pressure gradients~\cite{roma10}, fast ion dilution of the main ion species~\cite{tard08,holl11}, and Shafranov shift stabilization~\cite{bour05}. Here, we report on electromagnetic \textit{nonlinear} simulations of experimental discharges.

\begin{figure}[htbp]
	\centering
		\includegraphics[scale=0.5]{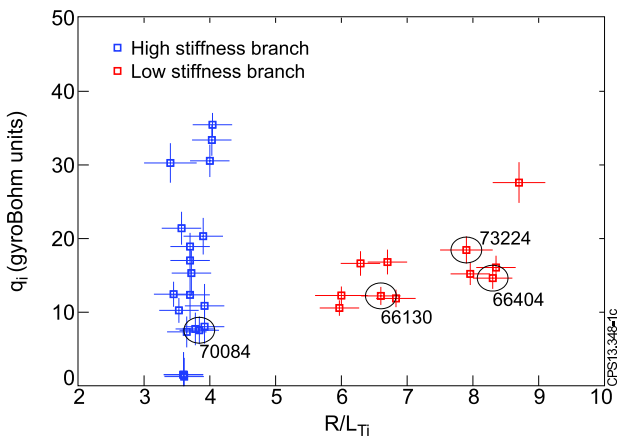}
		\caption{Ion heat flux versus logarithmic ion temperature gradients, from JET data presented in Ref.~\cite{mant11} showing a separation between high and low stiffness regimes at $\rho=0.33$. The specific discharges studied in this Letter are circled.} 
	\label{fig:figure1}
\end{figure}

\textit{Experimental discharges}.--A subset of discharges from the dataset described in Ref.~\cite{mant11} are analyzed at $\rho=0.33$, where $\rho$ is the normalized square root of the toroidal flux. The data splits into two branches corresponding to high and low stiffness, separated by heating scheme. While the discharges in both branches utilize ICRH, the discharges in the `low stiffness branch' also utilize significant NBI. We concentrate on the `low stiffness branch', and the specific discharges studied are circled in Fig.~\ref{fig:figure1}. The reduced stiffness is evident both from modulation experiments and from measured $R/L_{Ti}$ which are significantly greater than the modeled linear instability thresholds. We note however that the ion heat transport in discharge 66404 has been analyzed in Ref.~\cite{ryte11}, and the possibility of increased critical threshold also contributing to the observation is not ruled out. The correlation between reduced stiffness and low $\hat{s}$ is reported in Ref.~\cite{mant11}. 

Interpretative simulations of the selected discharges were carried out with the CRONOS~\cite{arta10} suite of integrated modeling codes. The fast ion profiles were calculated by NEMO/SPOT~\cite{schn11} for NBI-driven fast ions, and by SELFO~\cite{hedi02} for the ICRH-driven fast ions (${}^3$He). SELFO includes finite ion cyclotron orbit width effects, which is important for an accurate calculation of the ICRH fast ion pressure profile width. The interpretative analysis yields safety factor $q$ and $\hat{s}$ values within $\approx15\%$ of the MSE or polarimetry constrained EFIT calculations. The dimensionless parameters of the discharge fed into the nonlinear gyrokinetic calculations are summarized in Tab.~\ref{tab:summary1}. Details of the heating schemes for these discharges are in Ref.~\cite{mant09}. The ion heat flux and stiffness sensitivity to the various parameters were extensively studied. According to the simulations, the key parameters which impact the stiffness in this parameter regime are $\beta_e$ and the fast ion profiles. 

\begin{table}
\centering
\caption{\footnotesize Discharge dimensionless parameters. The $\hat{s}$ and $q$ values were calculated by CRONOS, assuming neoclassical current diffusion. The discharges were analyzed in a quasi-stationary state at flattop at $\rho=0.33$. The values are averaged over time $t=9.5-10.5$~s for discharges 70084 and 66130, over $t=6.5-7.5$~s for discharge 66404, and over $t=7-8$~s for 73224. For brevity, only the error bars for parameters which have significant impact on the system in the parameter range studied are displayed. The errors are statistical and do not take into account any possible systematic errors. $T_{\left(e,i\right)}$ are the electron and ion temperatures, $R/L_{\left(T,n\right)}$ the normalized logarithmic temperature and density gradients.   $\beta_e{\equiv}p_e/\left(B^2/2\mu_0\right)$, the ratio between the thermal electron and magnetic pressure. $\nu^*$ is the electron-ion collision frequency normalized to the trapped electron bounce frequency.}
\tabcolsep=0.09cm
\scalebox{0.84}{\begin{tabular}{c c c c c c c c c}
\label{tab:summary1}
Shot  & $\hat{s}$ & $q$ & $T_e/T_i$ & $R/L_{Ti}$ & $R/L_{Te}$ & $R/L_{ne}$ & $\beta_e [\%]$ & $\nu^*$ \\
\hline
70084 & $0.7$ & $1.7$ & $1.08\pm0.04$ & $3.5\pm0.5$ & $3.8$ & $1.4$ & $0.19\pm0.01$ & $0.07$ \\
66130 & $0.7$ & $1.8$ & $1.25\pm0.13$ & $6\pm0.4$   & $6.5$   & $2.4$   & $0.46\pm0.09$ & $0.04$ \\
66404 & $0.4$ & $1.8$ & $1.14\pm0.06$ & $8.6\pm0.9$ & $5.5$ & $3.8$ & $0.35\pm0.07$ & $0.02$ \\
73224 & $0.5$ & $1.7$ & $1.0\pm0.02$ & $9.3\pm1$ & $6.8$ & $1.3$ & $0.33\pm0.004$  & $0.038$ \\
\hline
\end{tabular}}
\end{table}

\textit{Simulation setup}.--The gyrokinetic turbulence code \textsc{Gene} was used in the radially local limit, justified since $1/\rho^*{\approx}500$ for the range of plasma parameters studied here~\cite{cand04,mcmi10}. $\rho^*$ is the ion Larmor radius normalized to the tokamak minor radius. Typical \textsc{Gene} grid parameters were as follows: perpendicular box sizes $[L_x,L_y]=[170,125]$ in units of $\rho_s{\equiv}c_s/\Omega_{ci}$, perpendicular grid discretizations $[n_x,n_y]=[192,48]$, $n_z=24$ points in the parallel direction, 32 points in the parallel velocity direction, and 8 magnetic moments. $c_s\equiv\left(T_e/m_i\right)^{1/2}$ and $\Omega_{ci}\equiv\left(eB/m_i\right)$. $x$ is the \textsc{Gene} radial coordinate, $z$ the coordinate along the field line, and $y$ the binormal coordinate. All simulations included kinetic electrons. Both an analytical circular geometry model~\cite{lapi09} as well as an experimental geometry were used. Extensive convergence tests were carried out for representative simulations throughout the parameter space spanned. 

The ion heat fluxes correspond to time-averaged values over the saturated state of the \textsc{Gene} simulations, and are in gyroBohm normalized units. The normalizing factor is $q_{iGB}=T_i^{2.5}n_im_i^{0.5}/e^2B^2R^2$, where $n_i$ is the ion density and $m_i$ the ion mass. However, for consistency with Refs.~\cite{mant09,mant11}, $n_e$ was used as a proxy for $n_i$ in the normalization in this work. For purely toroidal rotational flow shear, as assumed here, $\gamma_E{\equiv}\frac{r}{q}\frac{d\Omega}{dr}/(\frac{c_s}{R})$ is the normalized perpendicular flow shear rate. In the electromagnetic simulations, only the ${\delta}B_\bot$ fluctuations were computed, justified by the relatively low $\beta_e$ values. Including ${\delta}B_\parallel$ in the system had a negligible impact on the heat flux predictions.

\begin{figure}[htbp]
	\centering
		\includegraphics[scale=0.5]{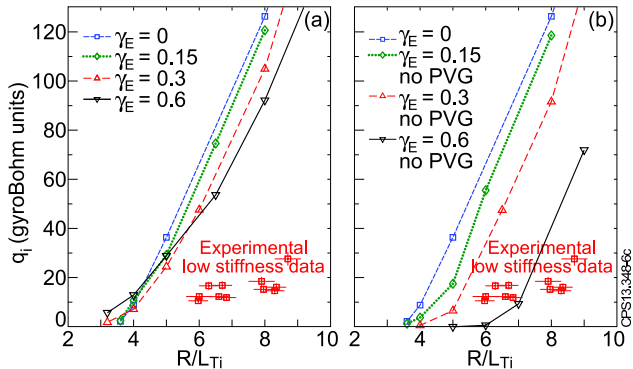}
		\caption{Ion heat flux in nonlinear \textsc{Gene} $R/L_{Ti}$ and $\gamma_E$ scans based on discharge 70084 parameters at $\rho=0.33$ ($q/\epsilon=11.8$ for circular geometry). $\hat{s}/q=0.2/1.3$ throughout. Results are shown both including (a) and neglecting (b) PVG destabilization. All runs were electrostatic, collisionless, used circular geometry, and assumed $T_e/T_i=1$. The results are compared with the low stiffness data at $\rho=0.33$ from Ref.~\cite{mant11}.}
	\label{fig:figure2}
\end{figure}

\textit{Impact of flow shear}.--The high and low stiffness branches are correlated with low ($\gamma_E\approx0.1$) and high ($0.3$) flow shear respectively. This correlation, together with concomitant low $\hat{s}$, was previously hypothesized to lead to the stiffness reduction. However, as shown in $R/L_{Ti}$ and $\gamma_E$ scans in Fig.~\ref{fig:figure2}, the modeled impact of the flow shear on the ion heat flux was not sufficient to reduce the flux level and stiffness to the experimental values. $\hat{s}=0.2$ in the scans, chosen to test the flow shear stabilization impact in the lower range of reasonable variations from the nominal value. Scans at the nominal $\hat{s}$ and $q$ values also did not reproduce the experimental observations.

The impact of parallel velocity gradient (PVG) destabilization is significant. This can be seen by comparing Fig.~\ref{fig:figure2}a, where the simulations included PVG, and Fig.~\ref{fig:figure2}b, which artificially excluded PVG. The significance of PVG drive has previously been seen in gyrokinetic simulations~\cite{kins05c}. Its impact decreases with decreasing geometrical factor $q/\epsilon$~\cite{high12}, where $\epsilon{\equiv}r/R$ is the inverse tokamak aspect ratio at the local radius $r$. For the circular geometry applied, $q/\epsilon=11.8$ at $\rho=0.33$. Additional simulations were carried out with exact experimental geometry, where $q/\langle\epsilon\rangle=10$ and $\langle\cdot\rangle$ denotes a flux surface average. In these cases the impact of the PVG destabilization is only slightly diminished, maintaining the basic picture portrayed in Fig.~\ref{fig:figure2}.

\begin{figure}[htbp]
	\centering
		\includegraphics[scale=0.5]{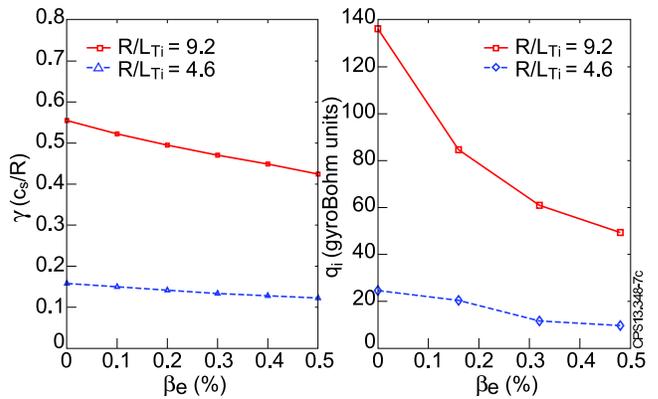}
		\caption{Linear growth rates (a) and nonlinear ion heat fluxes (b) calculated in $\beta_e$ and $R/L_{Ti}$ scans based on discharge 66404 parameters at $\rho=0.33$. In the linear scan, $k_y=0.4$ in units of $1/\rho_s$. Runs included collisions, experimental geometry, two species, and assumed $T_e/T_i=1$.} 
	\label{fig:figure3}
\end{figure}

\textit{Impact of electromagnetic effects}.--Here we present the significant impact of electromagnetic stabilization on the microturbulence in the discharges studied. Linear and nonlinear $\beta_e$ scans based on discharge 66404 parameters are shown in Fig.~\ref{fig:figure3}. The range of experimental $\beta_e$ values ($0-0.5\%$) lies significantly below the simulated kinetic ballooning mode (KBM) thresholds. Electromagnetic effects lead to linear ITG mode stabilization with increasing $\beta_e$~\cite{kim93}. For our parameters, this leads to a growth rate reduction of $\approx25\%$ at $\beta_e=0.5\%$, at the upper range of our experimental $\beta_e$ values. The degree of linear ITG mode stabilization, i.e., the relative reduction of $\gamma$ for $\beta_e>0$ compared with $\beta_e=0$, is stronger as $R/L_{Ti}$ is increased. This is consistent with the corresponding increase of the coupling between the electromagnetic shear Alfv{\'e}n wave and the ITG mode with pressure gradients at any given $\beta_e$ value~\cite{kim93}. 

A striking observation is that the \textit{nonlinear} electromagnetic ITG stabilization significantly exceeds the linear stabilization, increasing to $\approx65\%$ as compared with the linear $\approx25\%$ at the upper range of the experimental $\beta_e$ values. This is consistent with \textsc{Gene} results reported in Refs.~\cite{pues08,pues10,pues13b}, which correlated the enhanced nonlinear stabilization with increased relative zonal flow activity and zonal flow effective growth rates. This increase may be related to the predicted increased coupling to zonal flows in the electromagnetic regime~\cite{mili11}. Future work will investigate these dynamics further. 

\begin{figure}[tbp]
	\centering
		\includegraphics[scale=0.4]{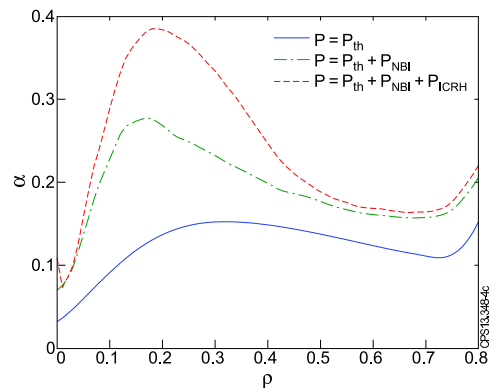}
		\caption{The comparative influence of thermal and suprathermal pressure components on $\alpha$ for discharge 66404. Increased $\alpha$ corresponds qualitatively to increased electromagnetic stabilization. The strong influence of suprathermal pressure for $\rho<0.4$ coincides with the measured low stiffness zone.}
	\label{fig:figure4}
\end{figure}

A key point is that the nonlinear electromagnetic stabilization can be significantly augmented by suprathermal pressure gradients. A parameter of merit for the strength of the electromagnetic impact on the linear ITG mode -- to which the nonlinear effect is likely linked -- is $\alpha{\equiv}q^2\sum_j\beta_j\left(R/L_{nj}+R/L_{Tj}\right)$, where $j$ sums over all particle species. $\alpha$ is a dimensionless measure of the pressure gradient. We stress that while not an exact parameterization in the general case, $\alpha$ nevertheless captures the qualitative dependency of the effect on the various relevant parameters~\cite{kim93,roma10}. For discharge 66404, the increase in $\alpha$ due to the modeled ICRH and NBI fast ion contributions is shown in Fig.~\ref{fig:figure4}. Importantly, the fast ions increase $\alpha$ while simultaneously not contributing to the ITG mode drive. The most significant fast ion contribution to $\alpha$ is at $\rho<0.4$, coinciding with the decreased stiffness zone in the experiments~\cite{ryte11}.



\begin{figure}[tbp]
	\centering
		\includegraphics[scale=0.5]{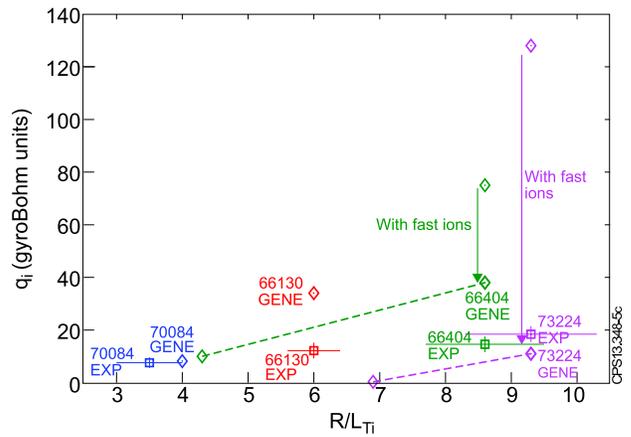}
		\caption{Comparison of nonlinear \textsc{Gene} simulations and experimental ion heat flux measurements for the five separate discharges at $\rho=0.33$. The importance of the fast ion contribution is underlined by the sensitivity studies carried out for discharges 66404 and 73224. The dashed lines connect the results of the nominal 66404 and 73224 simulations with results obtained at reduced $R/L_{Ti}$.}
	\label{fig:figure5}
\end{figure}

The importance of the suprathermal pressure in reproducing the experimental results is seen in Fig.~\ref{fig:figure5}. These simulations constitute the full comparison with the experiments and minimize the simplifying assumptions. Electromagnetic effects, collisions, flow shear, realistic $T_e/T_i$, impurities, fast ions, and experimental geometry are included. The fast particle populations induced by NBI and ICRH are treated as separate hot Maxwellian species, taking the average energy of the fast ion slowing-down distributions as the temperatures. For the range of discharges studied, $T_\mathrm{fast}=26-34$~keV for NBI accelerated D, and $T_\mathrm{fast}=20-23$~keV for ICRH accelerated $^3He$. 

For discharge 70084, agreement between the simulation and measurement was reached for input parameters (e.g., $R/L_{Ti})$ within the confidence intervals of the nominal values. Discharges 66130, 66404, and 73224 were all simulated with their nominal parameters. For 73224, agreement within $30\%$ of the experimental flux value was obtained. For discharges 66130 and 66404, agreement within a factor of 3 was obtained. When removing the fast ions, the ion heat flux for discharge 66404 was increased by a factor of $\approx2$, and for 73224 by an order of magnitude. The fast ion stabilization shown here is primarily an electromagnetic stabilization effect, providing significant flux reduction beyond ion dilution and the Shafranov shift linear stabilization. From dedicated simulations, we have also seen that the degree of stabilization does not depend on the relative values of $R/L_{T}$ and $R/L_{n}$ of the fast ion species, as long as the $\alpha$ value remains constant. 

The experimentally observed low stiffness is also captured by the simulations. This is indicated by reduced $R/L_{Ti}$ runs carried out for discharges 66404 and 73224, displayed in Fig.~\ref{fig:figure5}. The low stiffness for 73224 is accompanied by an enhanced threshold upshift, indicated by marginal stability at $R/L_{Ti}=6.9$, significantly above the linear threshold of $R/L_{Ti,\mathrm{crit}}\approx2.5$.  This is consistent with Ref.~\cite{pues10}, where a threshold shift was accompanied by a stiffness reduction when moving from the electrostatic limit to finite $\beta_e$. The seeming lack of threshold modification for discharge 66404 is attributed to residual activity of trapped electron modes, destabilized by the higher $R/L_n$ and observed at low $R/L_{Ti}$ in linear analysis of this discharge.


The remaining discrepancies in the flux values between the various simulations and measurements can be reconciled by reasonable variations of the input parameters -- such as $R/L_{Ti}$, $T_e/T_i$, $\hat{s}$, $q$, and $Z_\mathrm{eff}$ -- within the experimental uncertainties. $Z_\mathrm{eff}\equiv\left(\sum{Z_j^2n_j}\right)/n_e$ is the effective ion charge. However, the discrepancies observed when \textit{not} including the fast ions in an electromagnetic framework are clearly outside this envelope. 


We note that discharges in the `high stiffness' branch were also investigated. The significantly lower thermal and suprathermal pressure gradients led to a much reduced impact on the ion heat flux and stiffness reduction compared with the `low stiffness branch'. This is consistent with the electromagnetic stabilization mechanism being primarily responsible for the splitting of the experimental data into two separate stiffness branches.

Finally, the impact of the electromagnetic stabilization is stronger at low $\hat{s}$. This is shown in Tab.~\ref{tab:summary2}. The simulations -- based on discharge 66404 -- used circular geometry with $q=1.7$. This $\hat{s}$ dependence of the electromagnetic stabilization is in qualitative agreement with the experimentally observed decreased stiffness at low $\hat{s}$.

\begin{table}
\small
\centering
\caption{\footnotesize \textsc{Gene} simulations based on discharge 66404 with collisions, circular geometry, two species, and assumed $T_e/T_i=1$. The uncertainty values reflect the ion heat flux fluctuations during the saturated state. The electromagnetic stabilization is stronger at low $\hat{s}$, as reflected by the `stabilization factor', which is the ratio between the electromagnetic and electrostatic ion heat fluxes.}
\tabcolsep=0.11cm
\scalebox{0.8}{\begin{tabular}{c c c c}
\label{tab:summary2}
\multirow{2}{*}{$\beta_e$ [\%]} & \multirow{2}{*}{$\hat{s}$} & \multirow{2}{*}{$q_i$ [gyroBohm units]} &  Stabilization\\
&  &    & factor \\
\hline
0 & 0.2 & 180$\pm$14  & \multirow{2}{*}{3.5}\\
0.32 & 0.2 & 52$\pm$11 & \\
\hline
0 & 0.45 & 230$\pm$14 & \multirow{2}{*}{2.6}\\
0.32 & 0.45 & 88$\pm$16 & \\
\hline
0 & 0.7 & 246$\pm$26 & \multirow{2}{*}{2.7} \\
0.32 & 0.7 & 90$\pm$30 &  \\
\hline
\end{tabular}}
\end{table}

\textit{Summary and implications}.--
Based on gyrokinetic simulations with the \textsc{Gene} code, nonlinear electromagnetic stabilization of ITG modes by both thermal and suprathermal pressure gradients is shown to be the key factor leading to a reduced ion temperature profile stiffness regime at JET. This mechanism provides a clear explanation for the observations, as opposed to the previously hypothesized mechanism of concomitant low magnetic shear and high rotational flow shear, which is shown to be insufficient to lead to significant stiffness reduction. The electromagnetic stabilization is also seen to be more effective at low magnetic shear, in line with the experimental trends. For these discharges, the nonlinear electromagnetic stabilization due to fast ions is significantly greater than linear fast ion stabilization processes such as Shafranov shift stabilization, ion dilution, and linear electromagnetic stabilization. This effect has striking consequences for burning plasma tokamak scenarios, where essentially for larger devices flow shear is expected to be low but the fast ion component from fusion-$\alpha$ particles will be significant. Evidence of such improved ion energy confinement in JET DT plasmas has been seen~\cite{shar08,test12}. Furthermore, the increased strength of the effect at low $\hat{s}$ indicates an improved energy confinement extrapolation for burning hybrid scenarios, which contain a large volume of low $\hat{s}$~\cite{ITER6}. This applies for DT hybrid scenarios at JET -- which may achieve improved energy confinement beyond what has been observed in DD discharges -- as well as for future burning plasma experiments such as ITER. Finally, in the JET stiffness experiments performed until now, flow shear and suprathermal pressure gradients were co-correlated. This calls for additional experiments to be devised, on various machines, to decouple these parameters and allow further detailed investigation of their respective impact on transport. 


\textit{Acknowledgements}.--This work, supported by the European Communities under the contract of Association between EURATOM/FOM, was carried out within the framework of the European Fusion Programme with financial support from NWO. The views and opinions expressed herein do not necessarily reflect those of the European Commission. This work is supported by NWO-RFBR Centre-of-Excellence on Fusion Physics and Technology (Grant Nr.~047.018.002). The authors would like to thank C. Angioni, H. Doerk, R. Dumont, D.R. Hatch, E. Highcock, F. Millitello, F. Ryter, A. Schekochihin, M. Schneider, J. Weiland, and E. Westerhof for stimulating discussions. Resources of HPC-FF in J{\"u}lich are gratefully acknowledged. The authors are grateful to D.R. Mikkelsen for aiding with computational resources at the National Research Scientific Computing Center, which is supported by the Office of Science of the U.S. Department of Energy under Contract Nr.~DE-AC02-05CH11231. 

\bibliographystyle{unsrt}
\bibliography{PRLbib2}

\end{document}